\documentclass[12pt,a4paper,final]{iopart}

\usepackage[utf8]{inputenc}
\usepackage{iopams}  
\usepackage[T1]{fontenc}
\usepackage{textcomp}
\usepackage{microtype}
\usepackage{graphicx}
\usepackage{siunitx}
\usepackage{marvosym}
\usepackage{subcaption}
\usepackage{csquotes}
\usepackage[backend=bibtex, bibencoding=ascii, style=numeric, sorting=none, url=false, isbn=false, maxbibnames=99, firstinits=true, sortcites=true]{biblatex}
\AtEveryBibitem{\clearfield{note}} 
\renewbibmacro{in:}{%
  \ifentrytype{article}{}{\printtext{\bibstring{in}\intitlepunct}}} 
\usepackage[breaklinks=true,colorlinks=true,linkcolor=blue,urlcolor=blue,citecolor=blue]{hyperref}

\begin{document}

\title{Evolution of full phononic band gaps in periodic cellular structures}

\author{Maximilian Wormser$^1$\textdagger, Franziska Warmuth$^1$ and Carolin Körner$^2$}
\address{$^{1}$ Joint Institute of Advanced Materials and Processes (ZMP), 

University of Erlangen-Nuremberg, Dr.-Mack-Straße 81, 90762 Fürth, Germany}
\address{$^2$ Chair of Materials Science and Engineering for Metals (WTM), 

University of Erlangen-Nuremberg, Martensstraße 5, 91058 Erlangen, Germany}

\ead{\textdagger \hspace{0.3em} maximilian.wormser@fau.de}




\begin{abstract}
Cellular materials not only show interesting static properties but can also be used to manipulate dynamic mechanical waves. In this contribution, the existence of phononic band gaps in periodic cellular structures is experimentally shown via sonic transmission experiment. Cellular structures with varying numbers of cells are excited by piezoceramic actuators and the transmitted waves are measured by piezoceramic sensors. The minimum number of cells necessary to form a clear band gap is determined. A rotation of the cells does not have an influence on the formation of the gap, indicating a complete phononic band gap. The experimental results are in good agreement with the numerically obtained dispersion relation.
\end{abstract}

\vspace{2pc}


\section{Introduction}

Metamaterials are artificially manufactured materials with counterintuitive properties usually not found in nature \cite{Lu09}. Besides mechanical metamaterials (e.g.,~negative Poisson's ratio \cite{Sch11b, War16, Nov16}) and optical metamaterials (e.g.,~negative index of refraction \cite{Che10a, She01}), there are metamaterials that exhibit novel properties regarding their interaction with acoustic waves. One subset of these metamaterials are phononic band gap materials in which the propagation of phonons is prohibited in one or more frequency ranges \cite{Kus94}. The phononic band gap (PBG) is an intriguing property that offers many possible applications, e.g.,~sonic insulation \cite{Pen10, Ols09}, vibration control \cite{Jav16, Yan15} or acoustic wave guides \cite{Miy05, Ols09}. Since their initial discovery in 1992 \cite{Sig92} PBGs have therefore attracted a lot of interest in the scientific community. 



There are different groups that numerically describe PBGs in periodic structures with unit cells, usually applying periodic Bloch-Floquet boundary conditions \cite{Tra16, Pha06, Che13}. In contrast, the number of publications showing experimental verifications of phononic band gaps in three-dimensional structures is small. Typically, shakers and accelerometers are used as means of excitation and sensing of the acoustic waves in a sonic transmission experiment \cite{Sha14, Yan15, Tra15}. These methods are not suited for high frequency ultrasonic sound waves, though. Piezoceramics are a viable alternative for these experiments, since they work in the ultrasonic as well as in the audible range. Additionally, they can function as sensor and actuator \cite{Par06}. Numerical results already showed the existence of PBGs for a specific eigenmode of a strut-based cubic unit cell \cite{Lie14, War15}. In contrast to phononic crystals which usually consist of two distinct, periodically arranged materials with a mismatch of density and elastic constants \cite{Lu09}, the material presented in this paper consists of only one phase. In these materials, PBGs emerge from a sophisticated design of the cellular structure.

In the following, we present the experimental verification of the previously found numerically predicted PBGs in cellular structures. We show that PBGs exist in the presented cellular structures manufactured from Ti-6Al-4V powder by selective electron beam melting (SEBM). Furthermore, this study is to this date the first one to investigate experimentally how many cells are necessary to form a PBG. For this, samples with varying numbers of unit cells are examined in a sonic transmission experiment to identify the PBGs. Additionally, the influence of the orientation of the structures is investigated. Finally, the results are compared to a numerically determined dispersion relation.

\section{Methods}
\subsection{Sample preparation}
\begin{figure}
    \centering
    \includegraphics[width=0.9\textwidth]{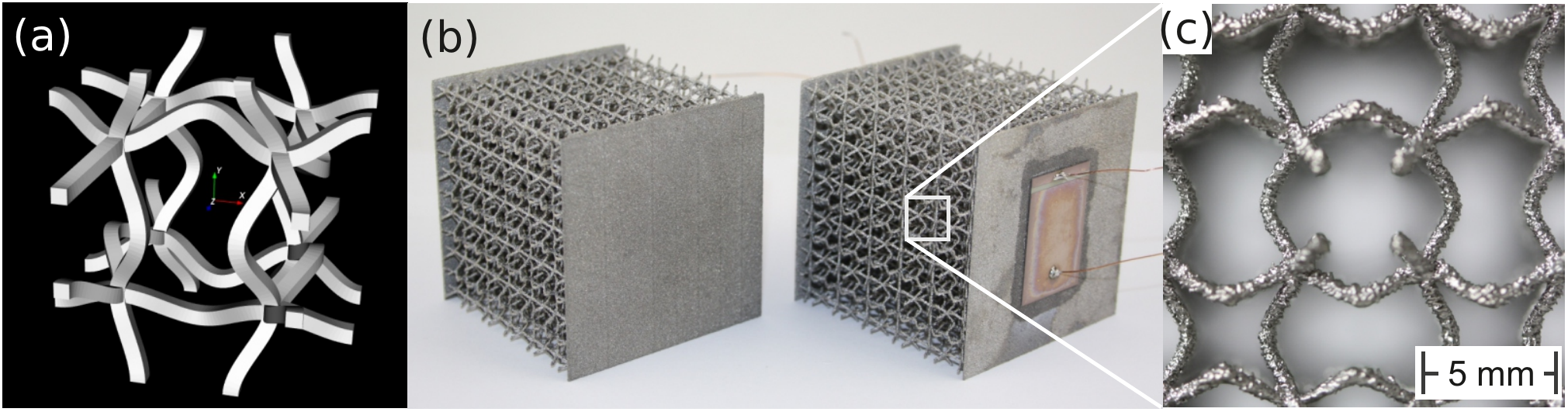}
    \caption{(a) CAD view of the unit cell which is the 61st eigenmode of a cubic cell with periodic boundary conditions. (b) Test samples made from Ti-6Al-4V by SEBM with applied piezoceramic actuator and sensor on the right sample. (c) Close-up view of the cellular structure.}
    \label{celloverview}
\end{figure}

The samples consist of a specific unit cell that was arranged using a CAD (computer aided design) software (see \hyperref[celloverview]{\autoref{celloverview}a}). This unit cell corresponds to the 61st eigenmode of the cubic unit cell with periodic boundary conditions \cite{Koer15, War15}. The node distance (i.e., where the struts intersect) is \SI{5}{\mm}, the amplitude of the struts is \SI{1}{\mm} and the strut thickness in the CAD file is \SI{0.25}{\milli\meter}, although the resulting strut thickness of the manufactured part will be mainly governed by the SEBM parameters. All samples have a width and height of \num{10 x 10} cells with varying number of 1, 3, 5, 7, 10, 15, 20 and 25 cells, where one cell is defined as the distance between two nodes. At two opposing sides, a thin wall covering the whole cross section is added to the structure to offer a flat surface for mounting the actuator and sensor by gluing. A finished sample and a close-up of the cellular structure can be seen in \hyperref[celloverview]{\autoref{celloverview}b,c}.

\begin{figure}
    \centering
    \includegraphics[width=0.6\textwidth]{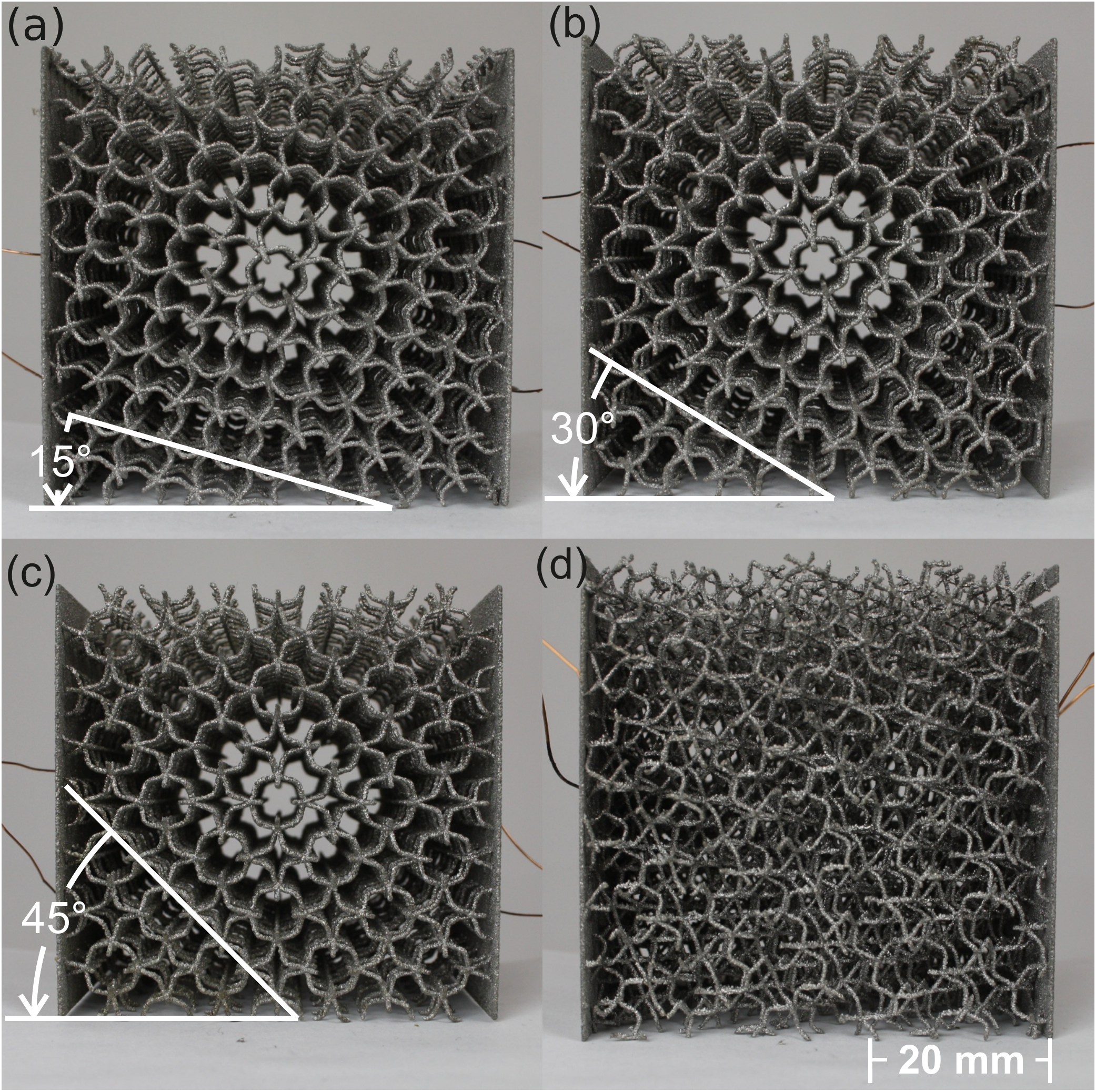}
    \caption{Photographs of samples with (a) \SI{15}{\degree}, (b) \SI{30}{\degree} and (c) \SI{45}{\degree} rotation around the length-axis of the sample. (d) A sample with rotation around all three axes.}
    \label{rotation_pics}
\end{figure}

Furthermore, samples with rotated unit cells (\SIlist{15; 30; 45}{\degree}; see \autoref{rotation_pics}) parallel to the side walls are built. Another sample with arbitrary rotation around all axes (\SIlist{85; 32; 58}{\degree} for $x$, $y$ and $z$, respectively) is constructed. The size of the rotated samples is identical to a \num{10 x 10 x 10} cells sample.

The structures are built layer by layer in the SEBM process using an \mbox{\textsc{Arcam AB Q10}} machine. Details on the process can be found in other publications (e.g., \cite{Koer16}) and will not be further discussed here. The material used in the process is Ti-6Al-4V. The process parameters of the SEBM process are identical for all samples used in this study (preheating temperature~$T$~=~\SI{730}{\celsius}, chamber~pressure~$p$~=~\SI{7e-6}{\bar}, beam~current~$I$~=~\SI{3}{\milli\ampere}, voltage~$U$~=~\SI{60}{\kilo\volt}, line~speed~$v$~=~\SI{450}{\milli\meter\per\second}, line~energy~$E$~=~\SI{0.4}{\joule\per\milli\meter}). 

On the flat surfaces (approx. \SIrange{400}{500}{\micro\meter} thick) on the two opposing sides of the samples one \SI{20}{\mm} $\times$ \SI{30}{\mm} piezoceramic modules (\textsc{PI Ceramic}) per side is applied using a thin layer of an acrylate based glue. The two electrodes per piezoceramic are contacted by soldering copper wire onto them.

The strut thickness was measured by computed tomography (CT) using a \textsc{Scanco MicroCT 40} for a single unit cell that was manufactured with the same parameters as the regular samples. It was calibrated using a  \SI{490}{\micro\meter} titanium wire. The measurement with a threshold of 750 was done at \SI{50}{\kilo\volt} voltage, \SI{160}{\micro\ampere} current and \SI{300}{\milli\second} exposure time with a voxel size of \SI{15}{\micro\meter} and an image size of \num{1024 x 1024} pixels. 

\subsection{Impedance and sonic transmission measurement setup}

\begin{figure}
\centering
\includegraphics[width=\textwidth]{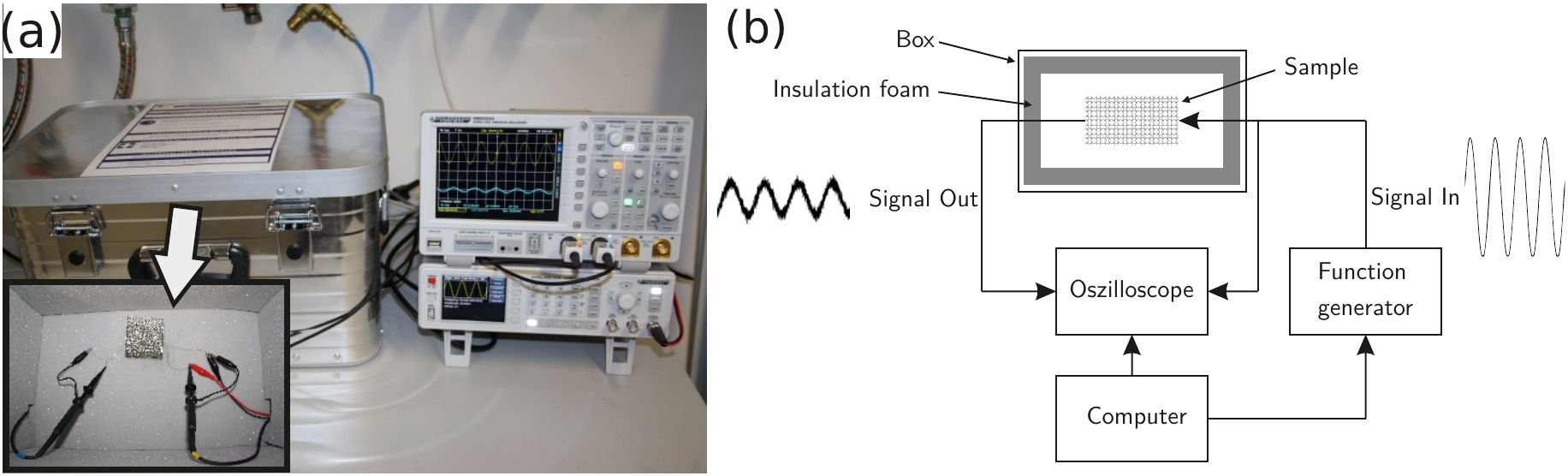}
\caption{(a) The sonic transmission setup with the oscilloscope and the function generator on the top right and bottom right, respectively. The inset shows how the samples are connected to the probes from the oscilloscope inside the of the box. (b) A schematic of the functional principle of the setup.}
\label{setup}
\end{figure}

To gain a better knowledge of possible influences of the piezoceramic modules on the transmission results we measure its impedance in the same frequency range where the sonic transmission is recorded. The impedance analysis is conducted on a \textsc{HP Impedance Analyzer 4194 A}. The impedance is measured from one electrode of a piezoceramic already applied to a sample to its other electrode.

The principle of the sonic transmission measurement setup (see \autoref{setup}) is to measure an incoming signal and compare it to the outgoing signal to see how much the transmitted signal is attenuated by the structure. Creating and analyzing the acoustic wave is done by the piezoceramic actuator and sensor, respectively. The actuator is connected to the function generator (\textsc{HAMEG HMF-2525}) which applies a sine wave function with a \SI{10}{\V} amplitude. The oscilloscope (\textsc{HAMEG HMO-2024}) probes the signal at the actuator and the signal at the sensor on the opposing side of the sample. This measurement is conducted in steps of \SI{1}{\kilo\Hz}, where each frequency is measured 64 times and automatically averaged by the oscilloscope. The computer automatically increments the frequency and saves the data using an in-house developed program. The sample is situated inside an aluminum box that is lined with insulating foam to prevent external influence from impacting the measurement. 

\subsection{Numerical calculation of the dispersion relation}

In order to numerically identify the PBGs of a periodic cellular structure, a dispersion relation has to be calculated. We determine the dispersion relation of the unit cell of \autoref{celloverview} using \textsc{COMSOL Multiphysics 5.1}. Periodic Bloch-Floquet boundary conditions are applied. A more detailed description of the method can be found in a previous publication \cite{War15}.

\section{Results and discussion}

\subsection{Impedance of the piezoceramic sensor/actuator module}

\begin{figure}
\centering
\includegraphics[width=0.9\textwidth]{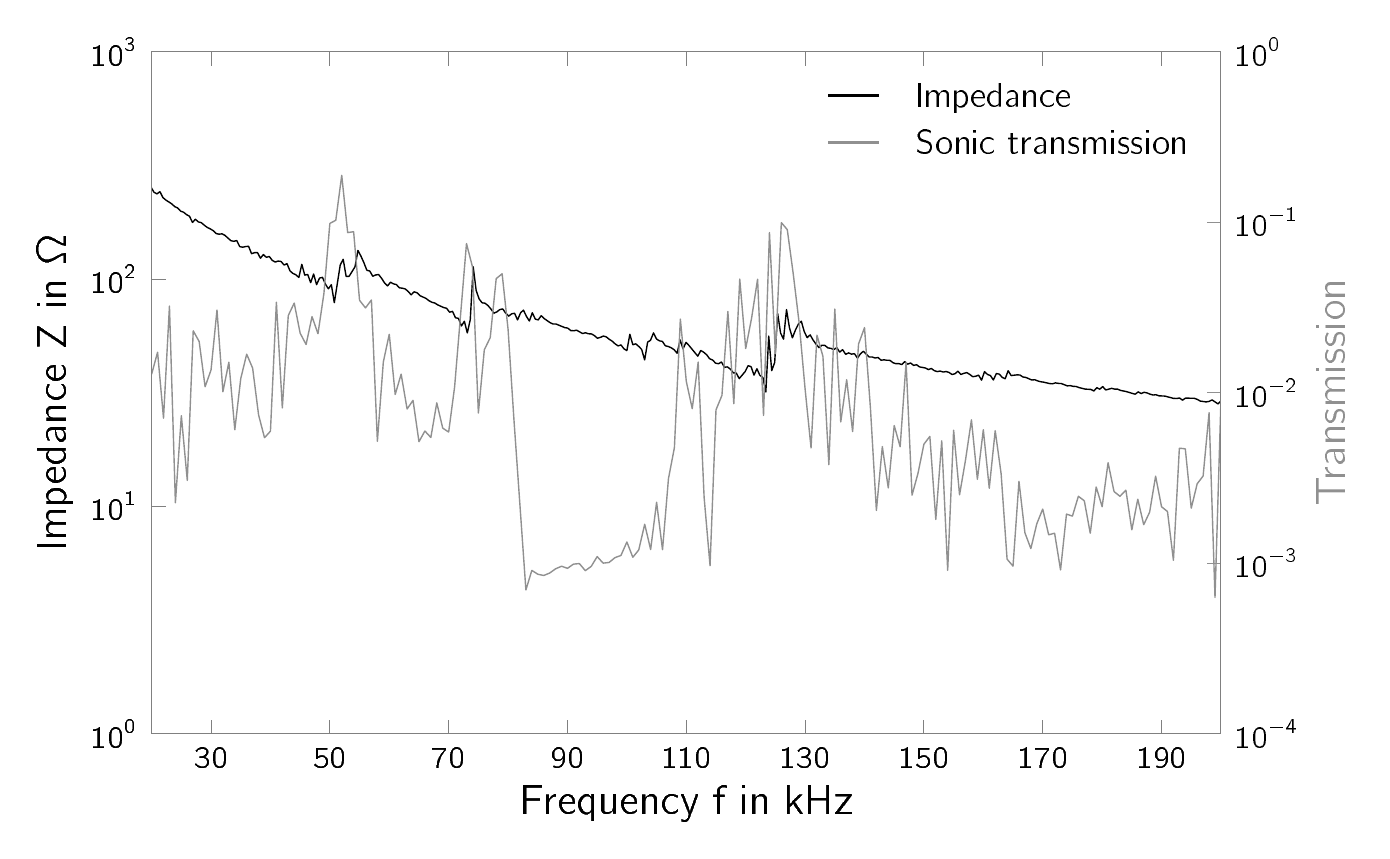}
\caption{A sonic transmission spectrum of the \num{10 x10 x10} sample (grey) and an impedance analysis of an actuator applied to the sample (black).}
\label{impedance}
\end{figure}

The impedance analysis of the piezoceramic modules mounted on the samples is depicted in \autoref{impedance}. Obviously, the piezoceramic actuators do not have a steady response behaviour. They show local maxima and minima at their \mbox{(anti-)resonance} frequencies which are located at around \SIlist{50; 75; 125}{\kilo\Hz}. A sonic transmission spectrum is plotted alongside the impedance curve to be able to qualitatively compare their shapes. The local minima in the impedance analysis coincide with the peaks of the sonic transmission spectrum. Conversely, the transmission gets lower where the impedance rises. All the other actuators show the same behaviour with the same characteristic resonance frequencies. In consequence, the elongation of the piezoceramic is stronger (weaker) at these \mbox{(anti-)resonance} frequencies. 

\subsection{Sonic transmission results for varying sample lengths}
\begin{figure}
\centering
\includegraphics[width=0.9\textwidth]{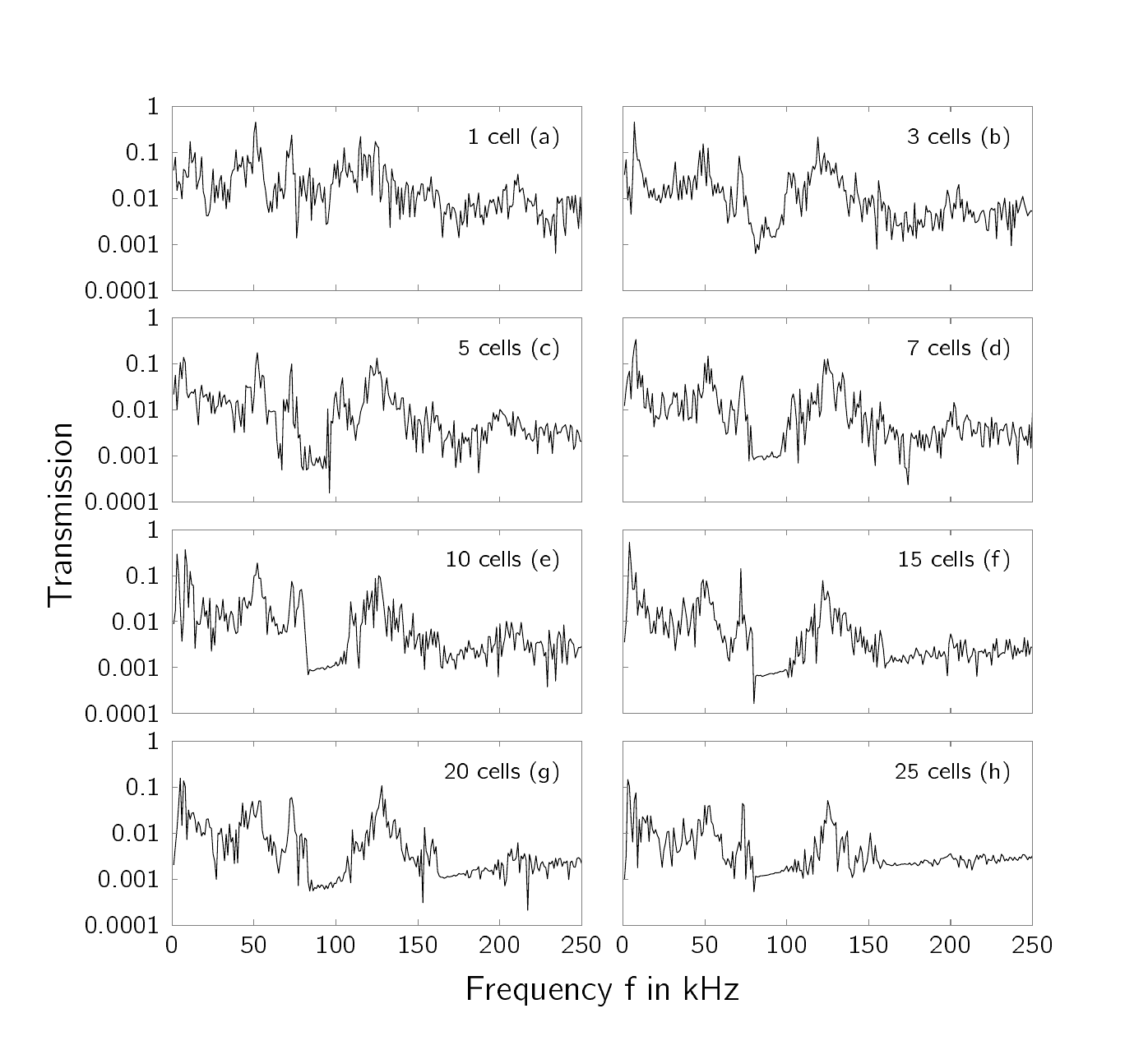}
\caption{Sonic transmission spectra for sample lengths of (a)--(h) 1, 3, 5, 7, 10, 15, 20 and 25 cells.}
\label{overview_L}
\end{figure}

Sonic transmission is characterized by the ratio of outgoing signal to incoming signal amplitude. This ratio is plotted over the frequency for sample lengths from 1 to 25 cells in \autoref{overview_L}. The lowest value is up to four orders of magnitude lower than the highest value, indicating the occurrence of strong signal attenuation or loss. The regions of interest, i.e., the phononic band gaps, are frequency ranges with consistently low sonic transmission ratios.

While there is no PBG visible in the short samples, a PBG clearly forms with longer samples. At a length of 7 cells and higher the PBG has a uniform minimum with low variation as opposed to the noisy signal on either side. The position of the PBG is roughly the same for all samples beginning at \SIrange{75}{80}{\kilo\hertz} and ending at \SIrange{100}{110}{\kilo\hertz}. The signal within the PBG is around one to two orders of magnitude smaller than the surrounding signals, usually at a transmission ratio of \num{0.001}. This also seems to be the resolution limit imposed by the accuracy of the oscilloscope.

For 15 cells or more a weak second gap evolves between \SI{160}{\kilo\hertz} and \SI{180}{\kilo\Hz}. While the gap is less pronounced here, the shape of the spectrum clearly indicates a region of interest. Especially the samples with a length of 20 and 25 cells show a smooth second PBG, comparable to the first PBG with at least 7 cells. The flat shape within the PBG is also observed in measurements with higher a resolution of \SI{0,1}{\kilo\hertz} instead of \SI{1}{\kilo\hertz}. These finer measurements of the non-PBG parts of the spectrum look even noisier while the PBG regions remain smooth.

From these results we can deduce that about 7 cells in transmission direction are necessary for a well-defined PBG to emerge. More than 7 cells up to 25 do not appear to be detrimental to the depth or width of the PBG. For the second gap to be visible the sample must have a length of at least 20 cells.

\subsection{Sonic transmission results for samples with varying orientation with respect to the active axis}
\begin{figure}
    \centering
    \includegraphics[width=0.9\textwidth]{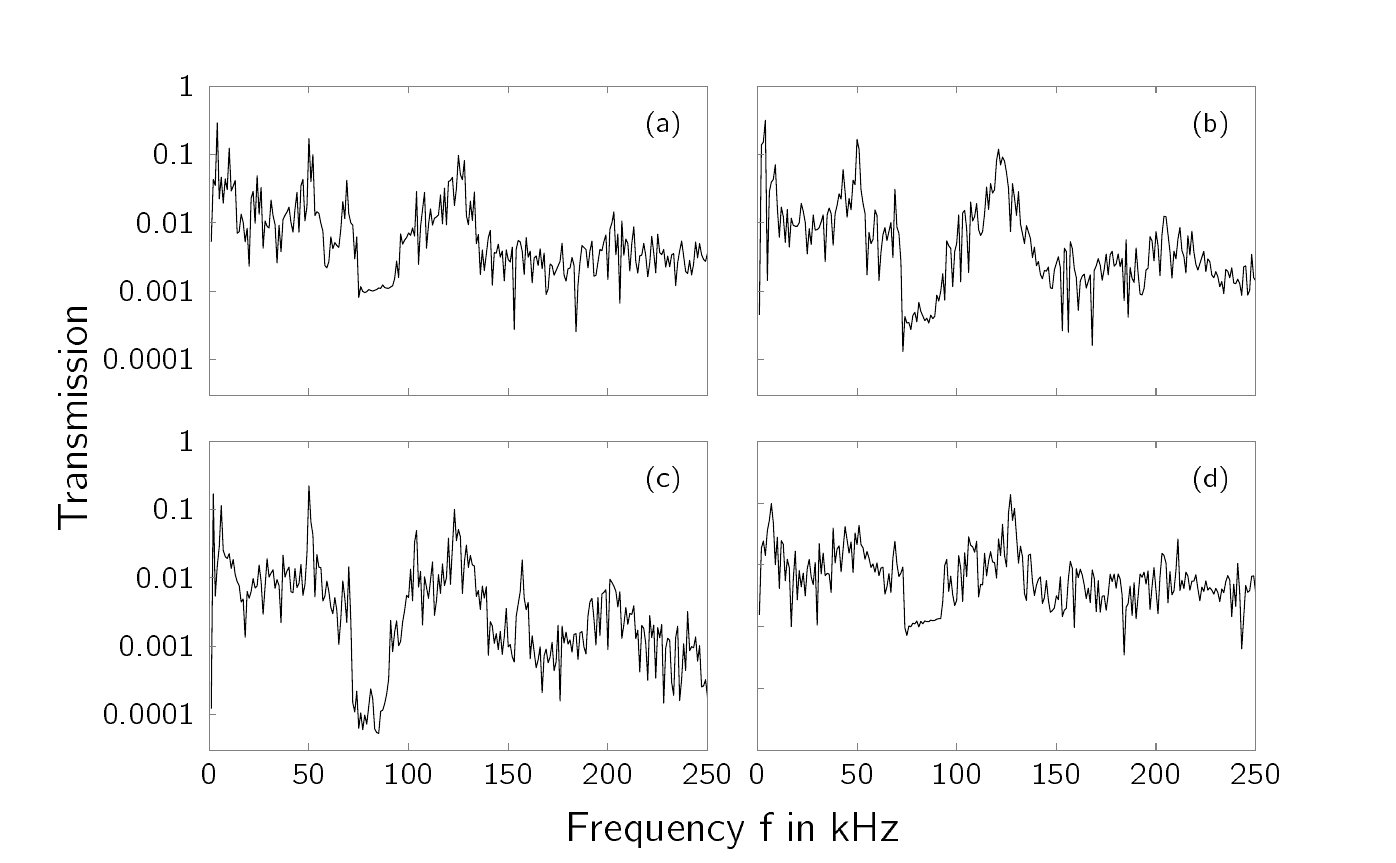}
    \caption{Sonic transmission spectra for samples with rotations of (a) \SI{15}{\degree}, (b) \SI{30}{\degree} and (c) \SI{45}{\degree} about the length of the sample and (d) an arbitrary rotation around all three axes.}
    \label{overview_rot}
\end{figure}

The spectra for the rotated structures (see \autoref{overview_rot}) qualitatively show the same results as the other samples. The PBG is visible and lies in the same range between \SI{75}{\kilo\hertz} and \SI{100}{\kilo\hertz}. Furthermore, the sample rotated around all three axes shows the same behaviour. This indicates a complete band gap, i.e., there is a PBG in all propagation directions of the mechanical wave.

There is no second PBG visible in any of the rotated samples. This is in line with the results of the last section, where a second PBG only emerged at a length of at least 20 cells, i.e., effectively twice the length of the rotated samples. As with the samples in the last section, the resonance frequencies of the piezoceramic actuator do not interfere with the PBGs.

\subsection{Comparison of experimental and numerical results}
In order to be able to compare the experimental data with numerical results, the thickness of the real samples has to be determined with a CT measurement of a unit cell. The CT measurement of the unit cell shows an average strut thickness of \SI[separate-uncertainty = true]{0,4905 +- 0,1165}{\micro\meter}. Given the large surface roughness and the fact that the roughness does not effectively contribute to the diameter of the strut in a mechanically relevant way \cite{Sua14}, the dispersion relation was calculated for a slightly thinner strut thickness of \SI{450}{\micro\meter}. 

\begin{figure}[ht]
    \centering
    \begin{minipage}[t]{0.75\textwidth}
    \centering \includegraphics[width=\textwidth]{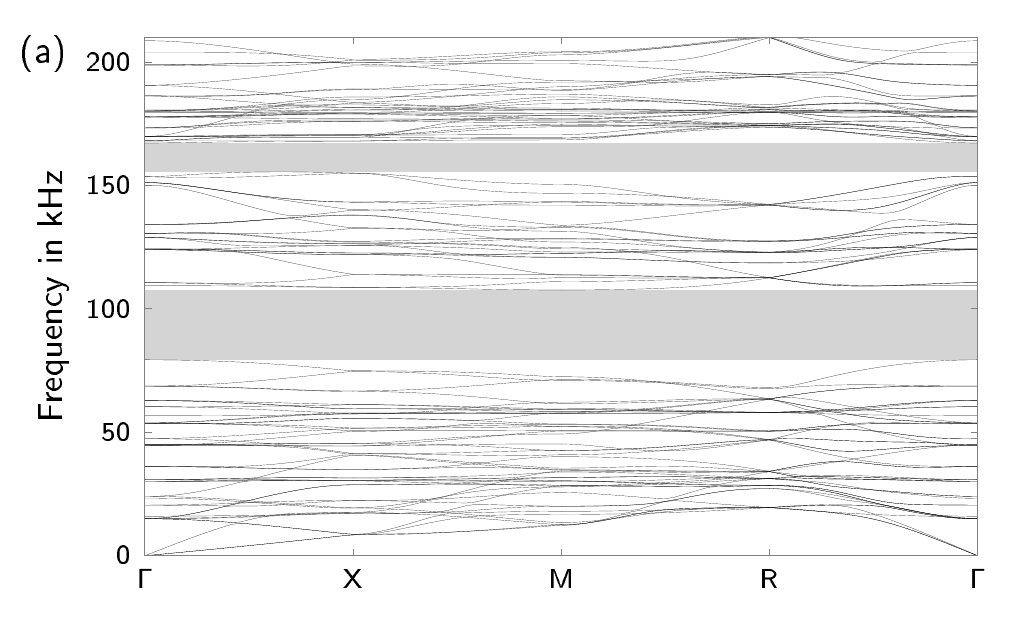}
    \end{minipage}
    \hspace{0.25cm}
    \begin{minipage}[t]{0.15\textwidth}
    \centering \includegraphics[width=\textwidth]{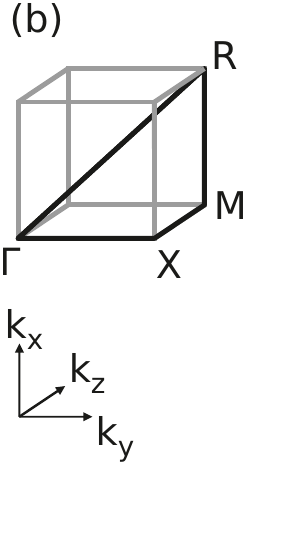}
    \end{minipage}
    
    \caption{(a) Dispersion relation of a unit cell with a strut thickness of \SI{450}{\micro\meter} along the reciprocal lattice points $\Gamma$-X-M-R-$\Gamma$. Complete PBGs are marked in grey. (b) Unit cell of the reciprocal lattice.}
    \label{disprel}
\end{figure}

The dispersion relation of the unit cell is shown in \autoref{disprel}. It shows two pronounced PBGs in the ranges from \SIrange{79}{108}{\kilo\hertz} and \SIrange{156}{167}{\kilo\hertz}. This numerical result is in good agreement with the experimentally determined PBGs of the samples with varying lengths and orientations, given that an exact determination of starting and ending values of the experimental PBG is not possible because a cut-off criterion would have to be defined arbitrarily.

\section{Conclusion}
In this study we examined the sonic transmission properties of a periodic cellular \mbox{Ti-6Al-4V} structure built by additive manufacturing (SEBM). The experiment showed that about 7 unit cells in transmission direction are necessary for a pronounced phononic band gap to evolve. 

The PBG is still present if the number of cells increases or the structure is arbitrarily rotated. Longer samples with 20 cells or more showed a second phononic band gap. The lack of influence of the orientation of the sample is furthermore evidence for a complete PBG, as suggested by the numerical results. The experimentally determined positions and widths of the PBG are in good agreement with numerical results. 

Further research will investigate on the effects of different geometric parameters to achieve wider and lower frequency PBGs with the underlying unit cell. A combination of different unit cells might open up possibilities to superpose PBGs.

\ack We gratefully acknowledge funding by the German Research Council (DFG) which, within the framework of its ”Excellence Initiative”, supports the Cluster of Excellence ”Engineering of Advanced Materials” at the FAU Erlangen-Nürnberg. Furthermore, we would like to thank the Bavarian state for the funding of the Application Center VerTec. We also thank Manuel Weiß and Stefan Rupitsch of the Chair of Sensor Technology of the FAU Erlangen-Nürnberg for fruitful discussions on the measurement setup and for letting us use the impedance analyzer. 

\printbibliography

\end{document}